# Experimental verification of the "rainbow" trapping effect in plasmonic graded gratings

Qiaoqiang Gan[1], Yongkang Gao[1], Kyle Wagner[2], Dmitri V. Vezenov[2], Yujie J. Ding[1] and Filbert J. Bartoli[1]*

1. Center for Optical Technologies, Electrical and Computer Engineering Department, Lehigh University, Bethlehem, PA 18015, USA

2. Department of Chemistry, Lehigh University, Bethlehem, PA 18015, USA

*Email: fjb205@lehigh.edu

**We report the first experimental observation of trapped rainbow[1] in graded metallic gratings[2-4], designed to validate theoretical predictions for this new class of plasmonic structures. One-dimensional tapered gratings were fabricated and their surface dispersion properties tailored by varying the grating period and depth, whose dimensions were confirmed by atomic force microscopy. Reduced group velocities and the plasmonic bandgap were observed. Direct measurements on graded grating structures show that light of different wavelengths in the 500-700nm region is "trapped" at different positions along the grating, consistent with computer simulations, thus verifying the "rainbow" trapping effect. The trapped rainbow effect offers exciting pathways for optical information storage and optical delays in photonic circuits at ambient temperature.**

Since Ebbesen's report on extraordinary optical transmission (EOT) through plasmonic hole arrays was published in 1998[5], plasmonics and metamaterials[6] have become topics of great interest. Advances in nanofabrication and characterization techniques now permit the study of a class of novel phenomena and devices with unique properties, including plasmonic subwavelength waveguides[7,8], plasmonic-assisted light emitting diodes[9], photodetectors[10], solar cells[11-13], integrated planar collimators[14] and plasmonic photon sorters[15], etc. By varying the surface nanotopology, the optical properties of surface plasmon polaritons (SPPs) can be tailored via so-called Surface Dispersion Engineering[16-18].

Recent theoretical investigations of "trapped rainbow" storage of THz waves in metamaterials[1] and plasmonic graded structures[2] generated considerable interest in novel slow light applications. It was shown that tapered waveguides with a negative refractive index core[1] and graded metallic grating structures[2,3] were capable of slowing a broadband "rainbow" to a standstill. By scaling the feature size of the graded grating structures down to nm level, it was theoretically predicted that telecommunication waves and even visible waves can also be trapped[4, 19].

Surface plasmon polaritons are known inherently to exhibit slow light behavior, as their group velocity decreases significantly when the photonic band edge is approached. Our recent investigations on one-dimensional (1D) graded metallic gratings demonstrated that the surface dispersion properties can be tuned by varying the geometric parameters. The dispersion relations for graded gratings vary monotonically with position, so that <u>incoming waves at different wavelengths can be "trapped" at different positions along the propagation direction on the grating.</u> Photonic crystal nano-cavities with graded hole size have recently been shown to exhibit adiabatically reduced group velocities for photonic modes at telecommunication frequencies[20]. However, "rainbow" trapping has not yet been demonstrated experimentally in plasmonic structures, or in the visible regime. In the present paper, one-dimensional graded metallic grating structures, whose surface dispersion properties are tailored by tuning the grating period and depth, were fabricated, and "rainbow" trapping experimentally observed in the visible domain. We also experimentally confirm key features of the predicted SPP dispersion properties, including observation of the photonic bandgap and measurement of the decrease in group velocity $v_g$ as the zone boundary is approached. Such solid-state plasmonic slow-light structures that operate at room-temperature provide a new means to control the group velocity[21], with the potential for incorporation into light-based (photonic) logic circuits.

## SURFACE DISPERSION ENGINEERING OF 1D METALLIC GROOVE ARRAYS

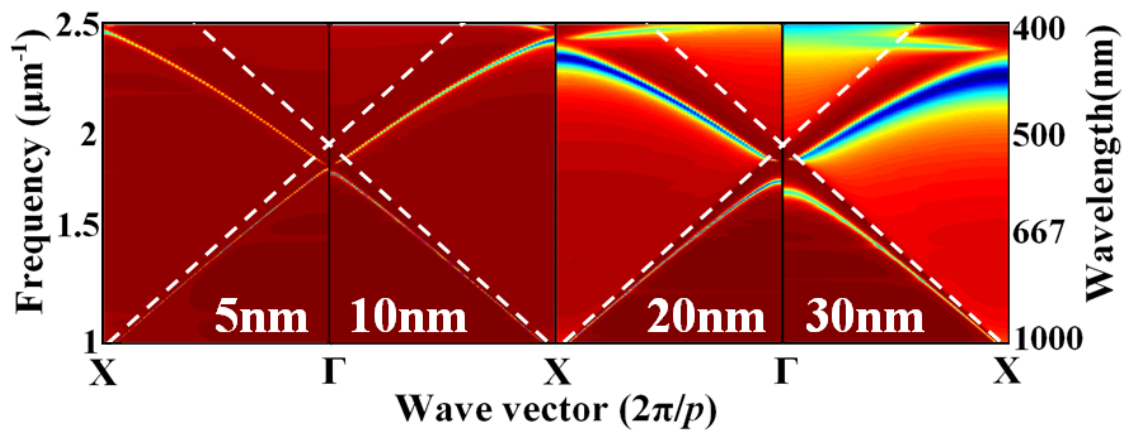

Figure 1  Dispersion curves of the nanopatterned silver gratings with groove depths of 5, 10, 20 and 30nm, calculated by the RCWA method. The dotted lines are light lines in free space.

It was theoretically predicted that the dispersion curves of 1D metallic grating structures, and consequently the $v_g$ of SPP modes, can be freely tuned by varying geometric parameters such as groove depth and period[4]. To demonstrate this tunability in the dispersion curves of nanopatterned surfaces, the angular-dependent reflection spectra of the silver grating surfaces were calculated using the rigorous coupled-wave analysis (RCWA) method[22], and their corresponding dispersion curves are plotted in Fig. 1. In these plots, the period and width of the grooves are 520nm and 150nm, respectively. One can see that the dispersion curve in the first Brillouin zone clearly bends downward, indicating a reduced $v_g$ for the SPP modes. Moreover, the group velocity reduction is greater for larger groove depths. Further modeling reveals how the shape of the dispersion curves can be tailored by control of the geometric parameters (see Fig. S1 in the supplementary material). Fig. 1 also shows that surface photonic bandgaps are created in these periodic nanopatterns, as was previously reported[16-18]. One can see that the bandgap occurs roughly in the 540-560nm wavelength region and broadens with groove depth. SPP modes in this frequency region cannot be supported by this nanopatterned surface. Such a groove structure has been suggested as a surface distributed Bragg reflector for a unidirectional surface wave coupler[23] or a bidirectional plasmonic splitter[24, 25].

## NANOSTRUCTURE FABRICATION AND OPTICAL CHARACTERIZATION

To confirm the theoretical predictions on the surface dispersion engineering, we fabricated 1D grating structures with various depths and periods using focus ion beam (FIB) milling (FEI Dual-Beam system 235). For the first experiment, eight gratings were fabricated on a 300-nm-thick layer of Ag evaporated onto flat fused silica microscope slides (Fisherbrand). The period and the width of this series of gratings were approximately 520nm and 150nm, respectively. Scanning electron microscope (SEM) images of two gratings with groove depths of 4.7 and 23.9nm are shown in Fig. 2. The eight structures have groove depths of 4.7, 6.8, 9.1, 11.7, 14.5, 16.7, 20.2 and 23.9 nm as measured by an atomic force microscope (AFM, Asylum Research MFP-3D) with an ultra-sharp tip (NanoAndMore USA Inc., SSS-NCH AFM probes).

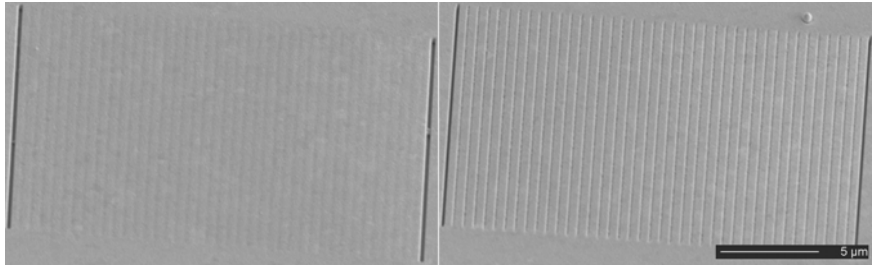

Figure 2 SEM images of two gratings with groove depths of 4.7 and 23.9 nm.

Next, we introduced two nanoslits to function as input and output couplers for a plasmonic Mach-Zehnder interferometer (MZI)[26], in order to characterize the optical properties of the surface grating structures. A schematic diagram of the plasmonic MZI is shown in Fig. 3 (a). The slit separation distance for our samples is approximate 21μm. Both the top surface (1: metal/glass interface) and bottom surface (2: air/metal interface) of the metal film can support SPP modes. When the SPP modes excited by the slit on the left propagate to the right slit, the SPP signals from the two optical branches (the top and bottom interfaces) interfere with each other, and thereafter modulate the far-field distribution of the scattered waves. The sensing arm of the plasmonic MZI is on the metal/air interface, whereas the reference arm is on the glass/metal interface. Measurement of the modulated far-field transmission as a function of wavelength permits us to probe changes in the effective refractive index and group velocity caused by changes in the grating structure. In a previous report, we successfully measured the spectral interference signals from such a structure[27]. In the present work, we employ 1D groove structures to engineer the surface dispersion properties of interface 2, modulate the interference patterns from the plasmonic MZI, and thereby measure the $v_g$ of the SPP modes supported by the nanopatterend surfaces. The far-field interference pattern is determined by

$$\cos\left[\frac{2\pi L}{\lambda}\left(n_{eff1} - n_{eff2}\right) + \varphi_0\right] \quad (1),$$

where $n_{eff}$ is the effective refractive index (ERI) of the metal/dielectric interface, $L$ is the slit separation distance, and $\varphi_0$ is an additional phase shift. For the flat interface 1, the ERI is given by the expression of $\sqrt{\frac{\varepsilon'_m(\lambda)n_2^2}{\varepsilon'_m(\lambda)+n_2^2}}$, where $\varepsilon'_m$ is the real part of the metal permittivity, and $n_2$ is the refractive index of the substrate. However, for the nanopatterned interface 1, $n_{eff1}$ increases with increasing groove depth. Eq. (1) predicts that the number of periods in the interference patterns per unit wavelength interval decreases and the interference peaks broaden when $n_{eff1}$ and $n_{eff2}$ approach each other[27]. The experimental results for the eight plasmonic MZIs are shown in Fig. 3 (b). In this plot, the low frequency background and high frequency noise have been numerically filtered using a Fast Fourier Transform technique (here we set the low frequency cutoff at ~4.344μm$^{-1}$ and the high frequency cutoff at ~101.358μm$^{-1}$). The data show that as the groove depth increases, the interference patterns broaden, and the number of the interference peaks per unit wavelength range decrease, which is consistent with the numerical modeling (see Fig. S2 in the supplementary material). Based on these interference patterns, the $v_g$ could be described approximately as:

$$v_{g2}(\omega) = 1 \Big/ \left[\frac{1}{v_{g1}(\omega)} - \frac{1}{L}\left(\frac{2\pi}{\Delta\omega}\right)\right] \quad (2),$$

where $v_{g1}$ is the group velocity of the SPP modes at the flat interface 1 (i.e. Ag/glass interface) and $v_{g2}$ is the group velocity of the SPP modes at the nanopatterned interface 2 (i.e. air/Ag interface), $L$ is slit separation distance. This expression should be valid as long as $v_g$ is not a rapidly varying function of the spectral oscillation period $\Delta\omega$. Under this condition, $v_{g2}(\omega)$ appears to be simply related to $\Delta\omega$ and $v_{g1}$[28] (see section 3 in supplementary material for additional detail). By employing the optical constants of Ag[29] to calculate the values of $v_{g1}$ at flat metal surfaces, the group index $n_{gr}$ or $v_{g2}$ can be estimated using Eq. (2). For clarity, extracted data for three depths are shown in Fig. 3 (c). One can see that $v_{g2}$ decreases gradually as the depth increases. However, it should be noted that the extracted group indices in Fig.3 (c) do not exceed 1.5, and the reduction in $v_g$ is not large. Use of this approximate approach to determine $v_g$ is not valid at the band edge where $v_g$ is too rapidly varying. From Fig. 1, one can see that the $v_g$ is predicted to decrease rapidly within a very small wavelength range just below the band edge. For example, for the grating structure with a depth of 30nm in Fig. 1, the $v_g$ decreases to approximately $0.55c$ at approximately 600nm and to $0.15c$ in the wavelength range of 585~587nm. The interference patterns in Fig. 3 (b) should be unresolvable as the band edge is approached, and other techniques are required to measure significantly reduced $v_g$ at the photonic band edge[30].

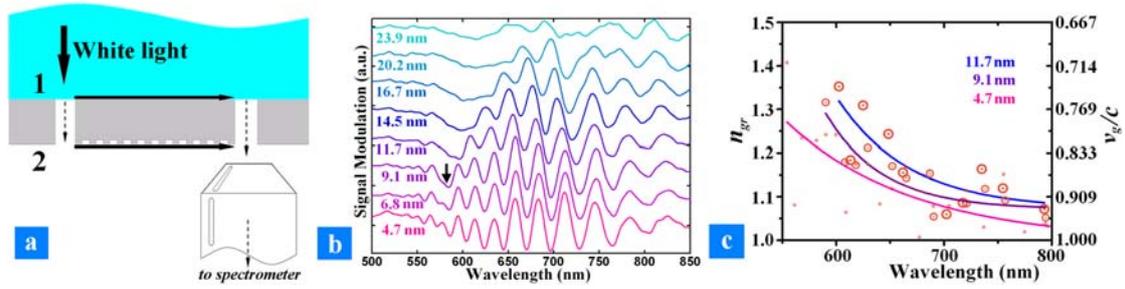

Fig. 3 (a) Illustration of the Plasmonic MZI measurement. (b) SPP-mediated spectral interference introduced by the SPP modes from the top and bottom surfaces: Here we study 8 doublet samples on a Ag film with groove depths of 4.7, 6.8, 9.1, 11.7, 14.5, 16.7, 20.2 and 23.9nm. (c) Extraction of the group index from 3 series of interference patterns in (b) using Eq. (2). The solid lines are exponential decay fit to the measured data to guide the eyes.

Plasmonic MZI measurements may be used to observe not only changes in the interference patterns due to a changing group velocity, but also the photonic bandgap. As shown in Fig. 3 (b), the interference pattern that is observed in the 570 to 580nm wavelength region for the shallow depth of 4.7nm becomes weaker in amplitude as the depth is increased to 6.9nm and 9.1nm, and eventually disappears for deeper grooves (as shown by the black arrows). Meanwhile, interference patterns at shorter or longer wavelengths are still observable, reflecting that a bandgap is formed in this wavelength region, consistent with the finite-difference time-domain (FDTD) modeling (see Fig. S2 in the supplementary material). With the deeper grooves, one can see that more and more interference patterns disappear, reflecting that the photonic bandgap is broadened as the depth is increased. Note that the interference patterns in the short wavelength region above the bandgap are not observed for deeper grooves because the loss of the SPP modes at higher frequencies increases significantly as the dispersion curve bends down.

To further confirm the tunability of the SPP bandgap, another series of structures was fabricated. The period and the width of the grooves are approximately 635nm and 200nm, respectively. The milling times for three of the four structures are identical to three of those in Fig. 3 (b), with nominal depths of 5nm, 10nm and 15nm. The fourth structure is a doublet with no grooves. According to the RCWA modeling shown in Fig. 4 (a), a bandgap is formed in the 650-680nm wavelength range (see the white arrows). This prediction is in a good agreement with the measured interference patterns shown in Fig. 4 (b). One can see that the interference patterns between the wavelengths of 660nm and 700nm disappear as the groove depth increases (see the black arrows), indicating the formation of the bandgap. The interference patterns at shorter and longer wavelengths are still observed. There is a slight difference in wavelength between the measured and calculated bandgap, which may be attributed to random surface roughness, fabrication imperfections, and physical characterization uncertainty. Nevertheless, these measurements provide clear evidence that the surface dispersion properties can be tuned by tailoring the dimensions of the nanopatterned surface.

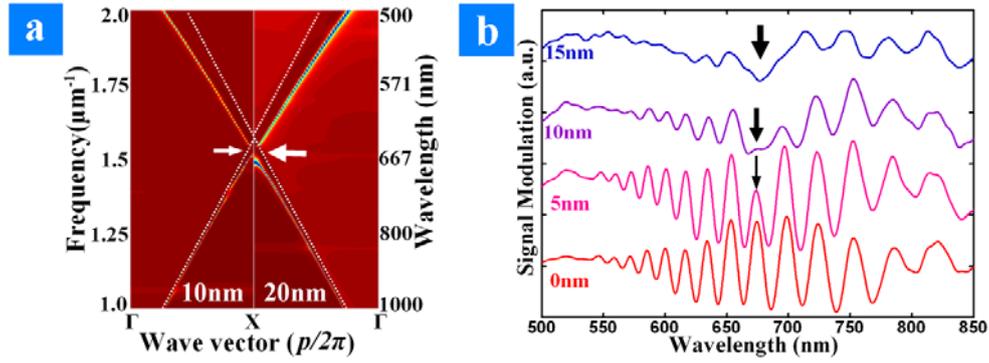

Figure 4 (a) The dispersion diagram of a surface groove structure with depths of 10nm and 15nm calculated by the RCWA method. The period and width of the grooves are 630nm and 150nm, respectively. The dotted lines are light lines in the free space. (b) Interference patterns from Plasmonic MZIs: Here we study 4 doublet samples on a Ag film with groove depths of approximate 0, 5, 10 and 15nm.

## METALLIC GRADED GRATINGS

The above observation of the bandgap and adiabatic variation of group velocity with changing groove depth is a clear illustration of tuning the surface dispersion properties of the nanograting structures. We now discuss the experimental verification of the "rainbow" trapping effect.

A graded grating, designed for the 500-700nm wavelength region, was fabricated by FIB milling, and the surface topology was measured by AFM. The period and the width of the grooves are approximately 475nm and 150nm, respectively. The milling time for each groove was increased steadily to achieve a grating with a linearly graded groove depth. According to the AFM characterization results, shown in Fig. 5 (a), the surface roughness at the bottom of the grooves is approximately 0.3nm, and the groove depth increases by approximately 1.5~1.8nm with each step. A nanoslit was fabricated at 13μm from the right edge of the graded grooves to function as an SPP coupler. It was illuminated from the bottom to couple free space light to SPP modes on the top surface, and light emanating from the nano-grooves can be observed from the top side. White light illumination is passed through a multi-band filter (Semrock, with transmission bands centered at 542nm and 639nm), and a colorful emission at red and green wavelengths is observed at different positions and shown in Fig. 5 (b), demonstrating the "rainbow" trapping effect. Here blue light would not be readily observed with this structure since 464nm is close to the SPP cutoff frequency and the propagation loss is predicted to be high. Subsequently, we used two bright line filters (Semrock) at 546nm (bandwidth 6nm) and 655nm (bandwidth 12nm) to control the illumination wavelengths. According to our theoretical predictions (results not shown here), the green light at 546nm and red light at 655nm should be trapped at the groove depths of approximately 30nm and 60nm, respectively [indicated by the triangles in Fig. 5(a)]. One can see from Fig. 5 (c) and (d) that the SPP emissions obviously extended to different positions along the graded grating structures. The central positions of the green and red emission fit reasonably well with the theoretical predictions [see triangles in Fig. 5 (b)].

The bright regions in these figures are relatively broad, possibly arising from: (1) the relatively broad transmission bands of the filters employed in the measurement; (2) increased scattering loss caused by surface roughness at the bottom of the nanogrooves; and (3) the breadth of the dispersion curve at the photonic band edge, and spread in values of group velocity in propagation loss, which is consistent with our previous numerical simulations on trapped "rainbow" in the visible region[4, 19]. All three reasons could contribute to the broad emission regions for different wavelengths. If the SPP modes are excited and guided from the opposite direction, they will reach the deepest groove first, which is approximately 100nm in depth. Theory predicts that the SPP modes should be scattered at once without propagating in the waveguide. To confirm this prediction, we fabricated another nanoslit positioned 13μm from the left edge of the graded grooves. This nanoslit couples the incident light to SPP modes and guides them in the opposite direction. It is obvious from Fig. 5 (e)-(g) that the emission scattered at the deepest grooves without further extension compared with the images shown in Fig. 5 (b)-(d).

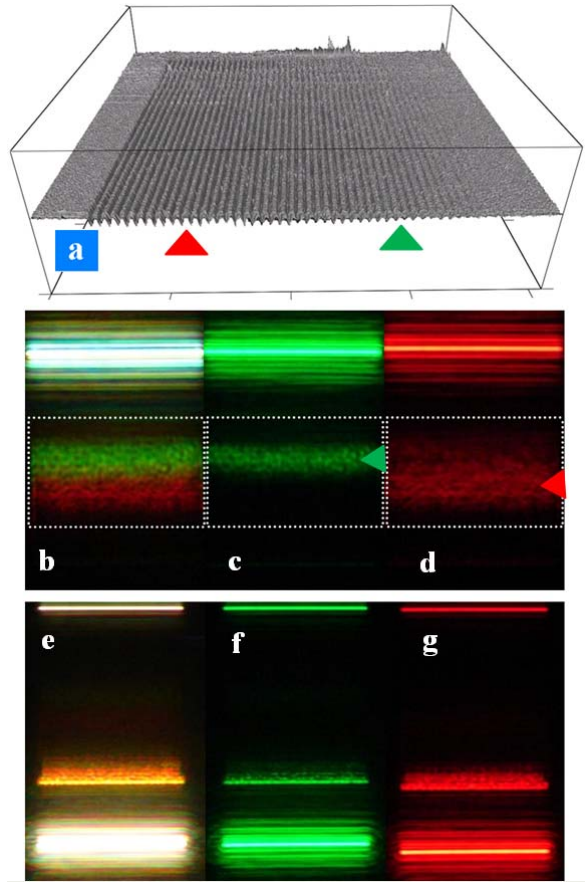

Fig. 5 (a) The AFM image of a graded grating. The period and the width of the grooves are approximately 475nm and 150nm, respectively. The groove depth increases from approximate 6nm to 100nm. (b)-(d) are images of the emission from the structure with different filters, respectively. (e)-(g) are emission images with the SPP modes coupled and propagating in the opposite direction in contrast to (b)-(d).

It is interesting that by varying the grade of the grating depths, the trapped positions for light of different wavelengths can be tuned freely, according to our previous simulations[2]. To validate this prediction, we fabricated another graded grating structure, whose period and width are fixed at 475nm and 150nm, respectively. The milling time for each groove was decreased by a factor of two while the number of the grooves was doubled. According to AFM characterization, the groove depth increased approximately 0.8~1.0nm by each step. The nanoslit input coupler was again fabricated 13μm from the edge of the graded grating. Fig. 6 shows the microscope images of the two graded grating structures side by side for comparison. One can see that as predicted the center positions of the green and red emissions are shifted slightly. In addition, the vertical widths of the SPP emission areas are broadened for the second structure with a smaller grade, which is consistent with the numerical modeling in ref. [2], demonstrating the tuning of the local dispersion properties of the graded gratings.

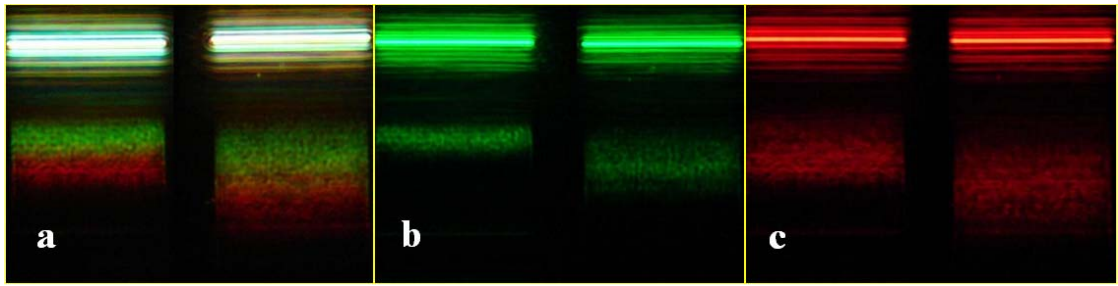

Fig. 6 Microscope images of graded gratings with different gradients. In this measurement, three filters are employed: (a) A multi-band filter (Semrock, transmission bands centered at 542nm and 639nm). (b) A green filter centered at 546nm (bandwidth 6nm). (c) A red filter centered at 655nm (bandwidth 12nm).

The concept of "trapped rainbow" storage of light[1-4] offers the potential to simultaneously slow down multiwavelengths in a single structure over a broad range of wavelengths. Our results verify that that the surface dispersion can be engineered by changing the structure of nanopatterned metallic surfaces. By using a graded surface grating structure, different wavelengths could be "trapped" at different positions on this structure. Although fabrication error and roughness introduce uncertainty, recently developed nanofabrication techniques that produce ultrasmooth patterned metals can be utilized to further validate theoretical predictions for this effect. This should allow improvements in design and fabrication of nanopatterned metal surfaces and optimization of their optical properties[31]. These new 'metamaterials' offer the ability to store photons inside solid-state structures at room temperature, with potential application to novel nanophotonic components, integrated photovoltaic devices, and biosensors on a chip.

**Acknowledgments**

The authors would like to acknowledge the support of this research by NSF (Award # 0901324).

**Author contributions**

Q.G. & F.J.B. conceived and designed the experiments. Q.G. & Y.G. fabricated the structures and performed the optical characterization. K.W. & D.V.V. performed the AFM characterization. Q. G., Y. J. D. & F. J. B. analyzed the data and wrote the paper.

Supplementary Material:

Experimental verification of the "rainbow" trapping effect in plasmonic graded gratings

Qiaoqiang Gan[1], Yongkang Gao[1], Kyle Wagner[2], Dmitri V. Vezenov[2], Yujie J. Ding[1] and Filbert J. Bartoli[1]*

3. Center for Optical Technologies, Electrical and Computer Engineering Department, Lehigh University, Bethlehem, PA 18015, USA
4. Department of Chemistry, Lehigh University, Bethlehem, PA 18015, USA

*Email: fjb205@lehigh.edu

**(1) Surface dispersion engineering with various geometric parameters**

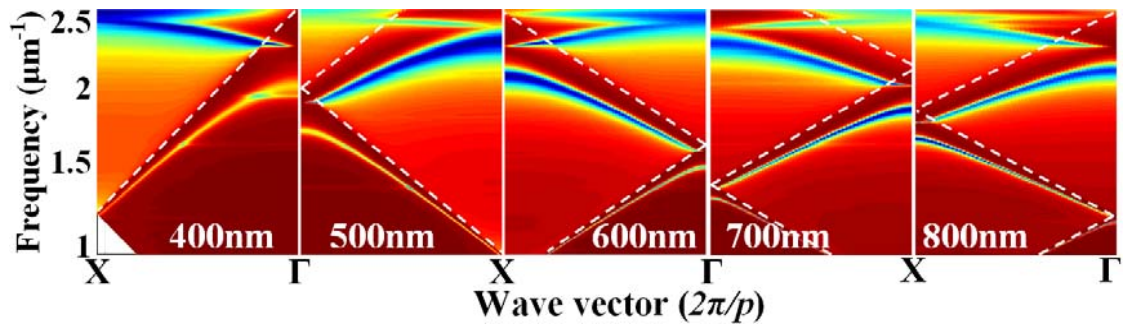

Figure S1 Surface dispersion engineering with various periods. The width and depth of the grooves are fixed at 150nm and 30nm, respectively. The period changes from 400nm to 800nm. One can see that the dispersion curves bend downwards relative to the dotted light lines and the frequency of the band gap can be tuned gradually by varying the period.

## (2) FDTD modeling of the interference patterns of the Plamsonic MZI with various groove depths

The spectral interference patterns of the plasmonic MZI were simulated using a two-dimensional (2D) FDTD model, as shown in Fig. S2. The thickness of the metal is 300nm, and the wavelength range considered is 500-850nm. The period and width of the grating structure are 520nm and150nm, respectively. The distance between the two slits is 21μm. One can see that as the depth of the grooves increases, the interference patterns are broadened gradually, and the number of the interference periods per unit wavelength range is gradually decreasing. This is qualitatively consistent with the measurement results in Fig. 2(b). In addition, as the depth increases, the interference patterns disappear in the wavelength range centered at ~550nm as indicated by the arrows in Fig. S2. It is in good agreement with the formation of band gap shown in in Fig. 1 (b) (calculated by the RCWA method). The slight quantitative difference between the FDTD numerical modeling and the measurement results may originate from the fabrication error and random surface roughness which cannot be fully incorporated into the FDTD simulations.

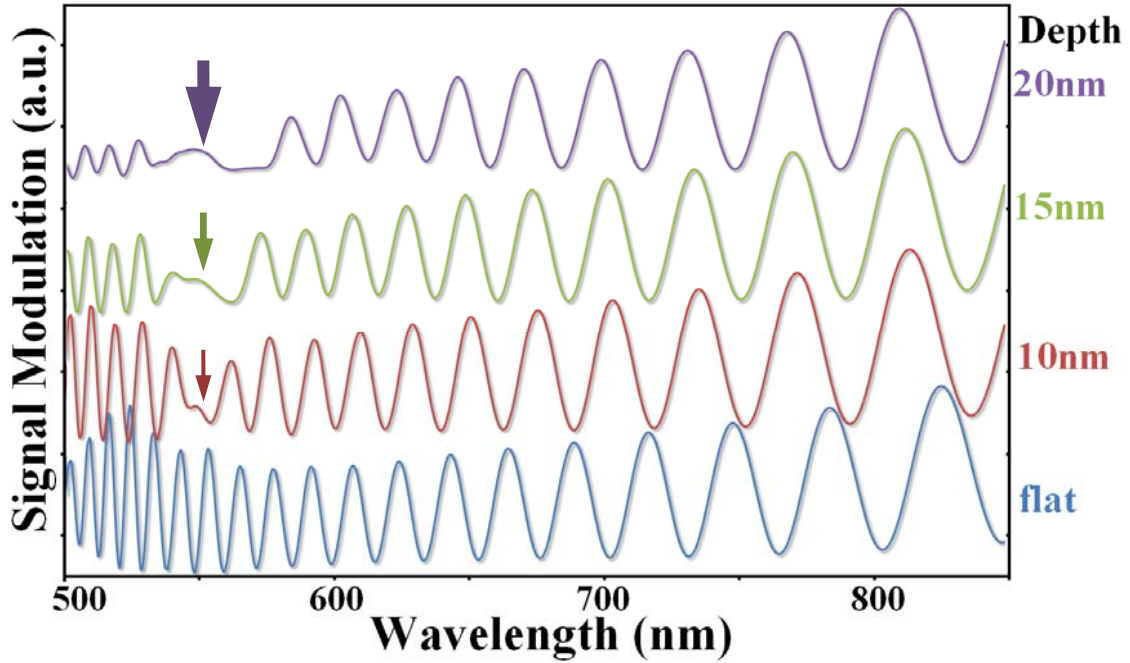

Figure S2 Spectral interference of the plasmonic MZI obtained using two dimensional FDTD simulations. Nonuniform mesh sizes are employed in this modeling: The edge grid sizes are Δx = 5nm and Δz = 2nm, and body grid sizes are Δx = 20nm and Δz = 20nm.

**(3) Extraction of the group velocity of the SPP modes at the nanopatterned interface**

According to Eq. (1), the phase factor of the interference pattern is given by

$$\varphi(\omega) = \{L[k_{sp1}(\omega) - k_{sp2}(\omega)] + \varphi_0(\omega)\}$$

(S1)

where $k_{sp1}(\omega)$ and $k_{sp2}(\omega)$ denote the wave vectors of the surface plasmon at the flat metal/glass interface 1 and nanopatterned interface 2, respectively. The derivative is

$$\frac{d\varphi(\omega)}{d\omega} = L\left[\frac{dk_{sp1}(\omega)}{d\omega} - \frac{dk_{sp2}(\omega)}{d\omega}\right] + \frac{d\varphi_0(\omega)}{d\omega} = L\left[\frac{1}{v_{g1}(\omega)} - \frac{1}{v_{g2}(\omega)}\right] + \frac{d\varphi_0(\omega)}{d\omega}$$

(S2).

Neglecting the additional phase shift term, i.e., $\varphi_0$ is assumed to be independent of wavelength, we obtain the expression for the group velocity of SPP modes at the nanopatterned interface 2:

$$v_{g2}(\omega) = \frac{1}{\dfrac{1}{v_{g1}(\omega)} - \dfrac{1}{L}\left[\dfrac{d\varphi(\omega)}{d\omega}\right]}$$

(S3)

Based on this equation, we employed an approximate procedure of counting minima and maxima of the interference patterns and obtained an expression for $v_{g2}$ [24], as shown in Eq. (2). The group velocity of the SPP modes at the nanopatterned interface 2, $v_{g2}$, appears to be simply related to the spectral oscillation period $\Delta\omega$ and $v_{g1}$. Using this equation, one can extract the group index $n_{gr}$ or $v_{g2}$ as shown in Fig. 3(c). It should be noted that the additional phase shift term, $\varphi_0(\omega)$, is also interesting and merits further investigations [S1].

Reference
[S1] G. Gay, O. Alloschery, B. Viaris de Lesegno, J. Weiner, & H. Lezec, Phys. Rev. Lett. **96,** 213901 (2006).